\documentclass[aps,preprint]{revtex4}

\usepackage{graphicx}

\begin{document}

\date{\today}

\title{Critical points in two-channel quantum systems} 

\author{
Hichem Eleuch$^{1,2}$\footnote{email: heleuch@physics.tamu.edu} and 
Ingrid Rotter$^{3}$\footnote{email: rotter@pks.mpg.de; corresponding author}}

\affiliation{
$^1$ Institute for Quantum Science and Engineering,
Texas A$\&$M University, College Station, Texas 77843, USA}
\affiliation{
	$^2$ Department of Applied Sciences and Mathematics, 
College of Arts and Sciences, Abu Dhabi University, Abu Dhabi, UAE}
\affiliation{
$^3$ Max Planck Institute for the Physics of Complex Systems,
D-01187 Dresden, Germany  }

\vspace*{1.5cm}

\begin{abstract}

Calculations for open quantum systems are performed usually by 
taking into account their embedding into one common environment,
which is mostly the common continuum of scattering wavefunctions.
Realistic quantum systems are coupled however often to more than 
one continuum. For example, the conductance of an open cavity needs
at least two environments, namely the input and the output channel.
In the present paper, we study generic features 
of the transfer of particles through  an open
quantum system coupled to two channels. We compare the results 
with those characteristic of a one-channel system. Of special 
interest is the parameter range which is influenced by singular 
points. Here, the states of the system are mixed 
via the environment. In the one-channel case, the resonance 
structure of the cross section is independent of the 
existence of singular points.
In the two-channel case, however, new effects appear caused by 
coherence. 
An example is the enhanced conductance of an open cavity 
in a certain finite parameter range. It is anti-correlated with 
the phase rigidity of the eigenfunctions of the 
non-Hermitian Hamilton operator.

\end{abstract}

\maketitle

\vspace{1cm}

\section{Introduction}
\label{intr}

Quantum systems are localized in a finite well-defined space area. 
This area may be determined by certain boundary conditions (e.g. 
in quantum dots or quantum billiards)
or by self-organization (e.g. in atomic nuclei). 
The shape of the nucleus is, indeed, an essential part of nuclear
spectroscopic investigations and is characteristic of every
nucleus. 

The spectroscopic properties of quantum systems are studied usually by
means of a Hermitian Hamilton operator $H$, the eigenstates of which
are discrete. Due to the embedding of the system into a continuum of
scattering wavefunctions, the lifetimes of 
its states become usually finite. This agrees with the 
experimentally well-known finite lifetime of the states of a quantum 
system which is described in the standard Hermitian quantum 
theory by means of the  tunneling mechanism. An exception are  
the well-known bound states in the continuum the lifetime of
which is infinitely long.  
These states occur under special conditions and are 
described well in different papers for different systems, 
e.g. \cite{wintgen1,wintgen2,rosabic,busabic}.

Moreover, the  embedding of a quantum system into a continuum of 
scattering wavefunctions causes another effect which is
not contained in  standard  Hermitian quantum theory:
since every state of the system is coupled to the continuum, all 
states of the system
may mix via the environment. This so-called  external
mixing (EM) of the states is a second-order process and
occurs additionally to the well-known
direct (internal) mixing of the states which is contained in every
Hermitian Hamiltonian $H$. In the standard Hermitian calculations, 
EM is effectively taken into account as an additional contribution to 
the internal mixing. Its characteristic features are lost
in this case.

Experimentally, an example of EM has been provided 
a few years ago in a mesoscopic system. It
has been shown in \cite{bird1}  that two distinct quantum
states  are coupled through a common continuum. 
In a further experiment, the authors were
able to show that EM  survives even under conditions of strongly
non-equilibrium transport in the system \cite{bird2}.

A possible EM of quantum states is very seldom considered in theory. 
It can easily be taken into account when the properties of the
system are described by a non-Hermitian Hamiltonian $\cal H$.
In such a case,  EM is involved explicitly in the eigenfunctions 
of $\cal H$, see  \cite{top,epj1,pra93,epj2,proj10}.

In \cite{proj10} the, at first glance, unexpected result has been
found that EM needs not to be considered when the system is embedded
in a common environment of scattering wavefunctions. This result
corresponds to the experience obtained from many numerical studies on
realistic systems in which EM is not at all considered.   Also
the influence of the singular exceptional points (EP) is not involved
in these  calculations.

Nevertheless, these results are not in contradiction with the
statement of non-Hermitian quantum physics that EPs and EM may
influence, to a great extent, the spectroscopic properties of open
quantum systems. The point is that EPs cause nonlinear processes in
their neighborhood \cite{top,epj1,pra93} which are able to compensate 
the contributions that arise from the EM of the states via the common
environment \cite{proj10}. Correspondingly, the 
cross section shows the same resonance structure in the one-channel 
case when calculated with EPs and EM or without them.

The situation is however different when the system is embedded 
not only in one common environment, but in several environments which
exist independently of one another. An example
of such a situation  is the decay of a
nuclear state into different states of the residual nucleus by
emission of a nucleon. 
In such a case, we have the different so-called partial widths 
that characterize the decay of the state $i$
of the system in one of the open  channels $c$. The different 
channels $c$ are defined by the
decay of the state $i$ into different states of the residual
nucleus by emission of a nucleon. 
A channel is open when the energy difference between the state $i$ and
the state of the residual nucleus is positive, i.e. when the
emission of a particle is allowed.

Another --and more important--
example is the transmission through, e.g., a quantum dot. In
this case, at least two different environments exist, namely that of
the entrance channel and that of the exit channel. Several numerical 
studies performed with a non-Hermitian Hamilton operator 
$\cal H$ on the basis of 
the tight-binding model have shown interesting non-trivial results
\cite{burosa1,burosa2}: 
the transmission probability is anti-correlated with the phase rigidity
of the eigenfunctions of the non-Hermitian Hamilton operator
(when averaged over energy in a certain energy window). That
means it depends on internal properties of the eigenfunctions 
of $\cal H$.

An enhancement of transmission through quantum dots or quantum
billiards is found also in other studies, mostly by varying the
coupling strength of the system to the environment. 
In \cite{naz1,naz2}, complex scaling is used while in 
\cite{cel1,cel2} the formation of a so-called superradiant state
is considered. In \cite{burosa1,burosa2,naz1,naz2}, 
the enhancement of the transmission is related to the properties of      
the eigenfunctions of the non-Hermitian Hamilton operator $\cal H$.

We mention here that the effects, which an environment has on the
transmission of particles 
through a quantum dot, have been considered also in, e.g., 
\cite{apollaro}. In this paper, the
competing effects of Markovian and non-Markovian mechanisms
have been investigated.

It is the aim of the present paper to find typical spectroscopic 
features of an open quantum system that is coupled to two 
particle-decay channels, 
which are independent of one another.
In Sect. \ref{eff} we discuss the effective non-Hermitian Hamiltonian 
$H^{\rm eff}$ which is used often in describing realistic systems.
In Sect. \ref{hamilt}, we sketch the formalism of the genuine
non-Hermitian Hamiltonian; and in Sect. \ref{num} we show some 
typical numerical results. A discussion of the obtained results 
can be found in  Sect. \ref{disc}. Their relation to the results of
other studies with non-Hermitian Hamiltonians is also discussed.
Concluding remarks can be found  in Sect. \ref{concl}.

\section{Effective non-Hermitian Hamilton operator $H^{\rm eff}$}
\label{eff}

Transfer and other processes with excitation of individual resonance
states are described successfully in non-Hermitian quantum physics
by using the effective Hamiltonian \cite{top}
\begin{eqnarray}
\label{heff}
H^{\rm eff} = H_0 + 
\sum_c V_{0c}\frac{1}{E^+ - H_c}V_{c0} \; .
\end{eqnarray}
Here, $H_0$ is the Hamiltonian describing the 
corresponding closed system with discrete states,
$(E^+ - H_c)^{-1}$ is the Green function in the 
continuum with the Hamiltonian $H_c$ describing the environment of
decay channels, and $V_{0c}, ~V_{c0}$ stand for the coupling of the
closed system to the different channels $c$ of the environment.
The non-Hermiticity of $H^{\rm eff}$ arises from the
second term of $H^{\rm eff}$, i.e. from the
perturbation of the  system occurring under the 
influence of its coupling  to the environment. 
It is complex: the real part arises from the principal value integral
and the imaginary one from the residuum (for details see \cite{top}).

Sometimes, $H^{\rm eff} = H_0 + \alpha  W $ is assumed 
where $W$ is imaginary, and the properties of the
system are studied as a function
of $\alpha$ (for examples see the
review \cite{top}). In \cite{sergi}  the non-Hermitian
operator is assumed to be $\hat {\cal H} = \hat H - i \hat \Gamma$. 
This operator allows us to study the general meaning 
of the imaginary
part of the non-Hermitian Hamiltonian, i.e. dissipation.
The information on the
considered physical system (embedded into a well-defined environment), 
which is involved in  $H^{\rm eff}$, is however lost.
In any case, the non-Hermiticity
of $H^{\rm eff}$ arises from the second term which is added  
to $H_0$ as a perturbation.

The eigenfunctions of a non-Hermitian Hamilton operator are biorthogonal
\begin{eqnarray}
\label{eif1}
{\cal H} |\Phi_i\rangle =  {\cal E}_i|\Phi_i\rangle \hspace*{1cm}
\langle \Psi_i|{\cal H} = {\cal E}_i \langle  \Psi_i|\; .
\end{eqnarray}
In the case of the symmetric 
Hamiltonian $H^{\rm eff}$, it is 
\begin{eqnarray}
\label{eif1a}
\Psi_i = \Phi_i^* 
\end{eqnarray}
and the eigenfunctions  should be normalized according to 
\begin{eqnarray}
\label{eif3}
\langle \Phi_i^*|\Phi_j\rangle = \delta_{ij} 
\end{eqnarray}
in order to smoothly describe the transition from a closed system 
with discrete states to a weakly open one with narrow resonance
states. As a consequence of (\ref{eif3}), the values of the 
standard expressions  are changed,
\begin{eqnarray}
\label{eif4}
 \langle\Phi_i|\Phi_i\rangle  =  
{\rm Re}~(\langle\Phi_i|\Phi_i\rangle) ~; \quad
A_i \equiv \langle\Phi_i|\Phi_i\rangle \ge 1 
\end{eqnarray}
\begin{eqnarray}
\label{eif5}
\nonumber
\langle\Phi_i|\Phi_{j\ne i}\rangle  = 
i ~{\rm Im}~(\langle\Phi_i|\Phi_{j \ne i}\rangle) =
-\langle\Phi_{j \ne i}|\Phi_i\rangle 
\end{eqnarray}
\begin{eqnarray} 
\label{eif6}
 |B_i^j|  \equiv |\langle \Phi_i | \Phi_{j \ne i}| ~\ge ~0 \; .  
\end{eqnarray}

The advantage to describe the properties of the open system by means of
$H^{\rm eff}$ consists, above all, in the possibility to use the results 
obtained for the corresponding closed system. In both cases,
the properties  are determined by
nothing but the eigenstates of the many-body system.
Usually, experimental results on inelastic scattering and transfer of 
particles through a small system are described well  
by $H^{\rm eff}$ (examples can be found in \cite{top}).
 
In (\ref{heff}), a singular point may appear, the so-called 
exceptional point (EP), at 
which two eigenvalues of $H^{\rm eff}$ coalesce \cite{kato}.
At these points, the two corresponding eigenfunctions are not 
orthogonal. Instead
\begin{eqnarray}
\label{sec8}
\Phi_1^{\rm cr} \to ~\pm ~i~\Phi_2^{\rm cr} \; ;
\quad \qquad \Phi_2^{\rm cr} \to
~\mp ~i~\Phi_1^{\rm cr}   
\end{eqnarray}  
according to analytical and numerical results 
\cite{ro01,magunov,gurosa,berggren,berggren2}. 
An EP is, according to its definition, related to the common 
environment in which the system is embedded. In other words, 
it is well defined under the condition that the system is 
embedded in only one continuum.

Far from EPs, the  coupling of the localized system to
the environment influences the spectroscopic properties of the system
only marginally \cite{top,proj10}. The influence is however 
nonvanishing also in this case, see e.g. the theoretical results
\cite{savin1} for very small coupling strength between system and 
environment, which  are proven experimentally 
\cite{savin2}. These
experimental results cannot be described by $H^{\rm  eff}$. 

Another deficit of  $H^{\rm eff}$ is that 
this formalism cannot be used for  the description of
systems with transfer of excitons.
An example is the photosynthesis in which not any eigenstates are
excited, see \cite{proj12}.

We will consider therefore in the next section \ref{hamilt} a genuine
non-Hermitian Hamilton operator $\cal H$
which is much less convenient for
numerical calculations than $H^{\rm eff}$. It gives us however 
a deeper insight into the reordering
processes occurring in open quantum systems under the influence of
the singular EPs.

\section{Genuine Non-Hermitian Hamiltonian $\cal H$}
\label{hamilt}

\subsection{Eigenvalues and eigenfunctions of ${\cal H}^{(2,1)}$}
\label{eig1ch}

To begin with, we sketch the features typical for an open quantum
system embedded in one common continuum. Details can be found in
\cite{top} and, above all, in \cite{proj10}. They can be discussed by
means of  the $2\times 2$ genuine non-Hermitian matrix
\begin{eqnarray}
{\cal H}^{(2,1)} = 
\left( \begin{array}{cc}
\varepsilon_{1}^{(1)} \equiv e_1^{(1)} + \frac{i}{2} \gamma_1^{(1)}  
& ~~~~\omega^{(1)}   \\
\omega^{(1)} & ~~~~\varepsilon_{2}^{(1)} \equiv e_2^{(1)} + 
\frac{i}{2} \gamma_2^{(1)}   \\
\end{array} \right) \; .
\label{ham2}
\end{eqnarray}
Here, the $e_i^{(1)}$ are the  energies of the localized states $i$
and the $\gamma_i^{(1)}$ are their widths \cite{comment1}.
The $\omega^{(1)}$ stand for the coupling matrix elements of the two
states via the common environment $(1)$. They  
are complex where Re($\omega^{(1)}$) arises from the principal value 
integral and Im($\omega^{(1)}$) from the residuum \cite{top}.
The complex eigenvalues ${\cal E}_i^{(1)} \equiv  E_i^{(1)} +
\frac{1}{2} \Gamma_i^{(1)}$ of ${\cal H}^{(2,1)}$ give the energies 
$E_i^{(1)}$ and widths $\Gamma_i^{(1)}$ of the states of the localized
part of the system.
We call the operator ${\cal H}^{(2,1)}$  genuine since it is not
related directly to any special quantum system (in contrast to  
(\ref{heff})). It contains nothing but two states characterized by 
their complex energies $\varepsilon_i$
and their coupling matrix elements $\omega$ via the environment. 

The eigenfunctions of $\cal H$ are biorthogonal, see 
(\ref{eif1}) to (\ref{eif6}). It is meaningful to define the
phase rigidity which is a quantitative measure for the biorthogonality 
of the eigenfunctions, 
\begin{eqnarray}
\label{eif7}
r_k ~\equiv ~\frac{\langle \Phi_k^* | \Phi_k \rangle}{\langle \Phi_k 
| \Phi_k \rangle} ~= ~A_k^{-1} \; . 
\end{eqnarray}
It is  smaller than 1. Far from an EP, 
 $r_k \approx 1$ while it approaches the value 
$r_k =0$ when an EP is approached.

Additionally to the Hamiltonian (\ref{ham2}), we will consider 
the non-Hermitian matrix 
\begin{eqnarray}
\label{ham0}
{\cal H}_0^{(2,1)} = 
\left( \begin{array}{cc}
\varepsilon_{1}^{(1)} \equiv e_1^{(1)} + \frac{i}{2} \gamma_1^{(1)}  & 0   \\
0 & ~~~~\varepsilon_{2}^{(1)} \equiv e_2^{(1)} + \frac{i}{2} \gamma_2^{(1)}   \\
\end{array} \right) 
\end{eqnarray}
which describes the system without any
mixing of its states via the environment. In other words, $\omega =0$ 
corresponds to vanishing EM of the eigenstates. The eigenfunctions
$\Phi_i$ of ${\cal H}^{(2,1)}$ can be represented in the set of
eigenfunctions  $\{\Phi_i^0\}$ of ${\cal H}_0^{(2,1)}$,
\begin{equation}
\label{eif12}
\Phi_i=\sum \, b_{ij} ~\Phi_j^0 \; ;
 \quad \quad b_{ij} = \langle \Phi_j^{0 *} | \Phi_i\rangle  
\end{equation}
under the condition that the $b_{ij}$ are normalized by 
$\sum_j (b_{ij})^2 = 1$. The coefficients $|b_{ij}|^2$ 
differ from the $(b_{ij})^2$. They contain the
information on the strength of EM. 

The main features characteristic of open quantum systems are described
well by the eigenvalues and eigenfunctions  of (\ref{ham2}).
Typical values related to the eigenfunctions are 
the phase rigidity (\ref{eif7}) and the contribution of  EM
(\ref{eif12}) to their purity. All these values contain the 
influence of the environment. They are proven experimentally, for 
details see \cite{top,proj10,ropp}.

\subsection{Schr\"odinger equation with ${\cal H}^{(2,1)}$}
\label{sour1}

The Schr\"odinger equation $({\cal H}^{(2,1)}
- {\cal{E}}_i^{(1)} |\Phi_i^{(1)}  \rangle =0$
may be rewritten into a Schr\"odinger equation with source term
\cite{top,proj10},
\begin{eqnarray}
\label{eif11}
({\cal H}^{(2,1)}_0  - {\cal E}_i^{(1)}) ~| \Phi_i^{(1)} \rangle  = -
\left(
\begin{array}{cc}
0 & \omega \\
\omega & 0 
\end{array} \right) |\Phi_i^{(1)} \rangle \; . 
\end{eqnarray}
In this representation, 
the coupling $\omega$ of the  states  $i$ and ${j\ne i}$ 
of the localized system
via the common environment of scattering wavefunctions (EM)
is contained in the source term of the Schr\"odinger equation, for
details see \cite{top}.

Far from EPs, the  coupling of the localized system to
the environment influences the spectroscopic properties of the system
only marginally \cite{top,proj10}. The influence is however 
nonvanishing also in this case, see e.g. the experimental results 
\cite{savin2} for the case that the coupling between system and environment 
is very small.

In the neighborhood of EPs, the coupling between 
system and environment causes --according to mathematical 
studies-- nonlinear effects in the Schr\"odinger equation 
(\ref{eif11}), see \cite{top,proj10}. 
Among others, these effects lead to a conservation of the resonance 
structure of the cross section in the one-channel case
which is therefore unaffected by EM and by the existence of EPs.   
Thus, the one-channel case cannot be used in order to test the 
results of the non-Hermitian formalism.

\subsection{Eigenvalues and eigenfunctions of ${\cal H}^{(2,2)}$}
\label{eig2ch}

Let us now consider the $4\times 4$ non-Hermitian matrix
\begin{eqnarray}
\label{ham22}
{\cal H}^{(2,2)} = 
\left( \begin{array}{cccc}
\varepsilon_{1}^{(1)} 
 &
~~~~\omega^{(1)} & 0 & 0  \\
\omega^{(1)} & ~~~~\varepsilon_{2}^{(1)}  
& 0 & 0 \\
0 & 0 & \varepsilon_{1}^{(2)}  
 &
~~~~\omega^{(2)}  \\
0 & 0 &
\omega^{(2)} & ~~~~\varepsilon_{2}^{(2)} 
  \\
\end{array} \right) \; .
\end{eqnarray}
Here, $\varepsilon_i^{(1)}\equiv e_i^{(1)} + \frac{i}{2}
\gamma_i^{(1)} $ and 
$\varepsilon_i^{(2)} \equiv e_i^{(2)} + \frac{i}{2} \gamma_i^{(2)}$
are the complex energies of the 
localized states $i$ relative to channel $1$
and $2$, respectively \cite{comment1}. 
Usually $\varepsilon_i^{(1)} \ne \varepsilon_i^{(2)}$.
The $\omega^{(1)}$ and $\omega^{(2)}$  stand for the coupling matrix 
elements of the two
states via the  environment $1$ and $2$, respectively.
The Hamiltonian (\ref{ham22}) includes the fact that the complex
energy $\varepsilon_i^{(c)}$ of the localized state
$i$ is different relative to the two different channels $c$. 

It might be astonishing that the Hamiltonian (\ref{ham22})
contains four states instead of the original two states. The point is 
the following. Two states may mix, independently of one another, via each of 
the two environments with the result that the energies $\varepsilon_i^{(c)}$
depend not only on
the state number $i$ but also on the channel $c$.
That means, every state is formally doubled, since it is embedded into  
two different environments (channels).
The two environments are different from and orthogonal to one another.  
Further, the two states with equal $i$ and different $c$ arise 
from the same state $i$ of the localized part of the system. The
zeros in the matrix  (\ref{ham22}) express the corresponding fact that 
the two states $i$ relative to the
two channels 1 and 2 
cannot interact with one another.  

The eigenvalues  ${\cal E}_i^{(c)} \equiv E_i^{(c)}
+\frac{i}{2}\Gamma_i^{(c)}$   and eigenfunctions $\Phi_i^{(c)}$ 
of (\ref{ham22}) are characterized also by two numbers: the number $i$  of 
the state ($i=1, 2$) of the localized part of the system and the
number $c$  of the channel ($c=1, 2$), called environment, in which the
system is embedded. Usually,  $E_i^{(1)} \ne E_i^{(2)}$ and 
$\Gamma_i^{(1)} \ne \Gamma_i^{(2)}$. 
Also the wave functions $\Phi_i^{(1)}$ and $\Phi_i^{(2)}$ 
differ from one another due to the EM of the 
eigenstates via the environment $1$ and $2$, respectively.
That means, the system described by (\ref{ham22}) has formally four
states (from a mathematical point of view). The original two states
related to the values $i=1$ and $i=2$ are doubled due to the fact that
each state $i$ is coupled to the two channels $c=1$ and $c=2$.

We mention further that the Hamiltonian (\ref{ham22}) is formally the
same as the Hamiltonian (24) or rather (1) in \cite{proj12}. There is
however a fundamental difference: in (1) in \cite{proj12}, the transition
of excitons (expressed by  fluctuations caused by EPs) is 
considered while (\ref{ham22}) describes the transition of
particles. Furthermore, in (1) in \cite{proj12} the whole system is
fully embedded into both environments (1) and (2) which both are
of completely different nature and exist independently of one
another. In contrast to this, the Hamiltonian   (\ref{ham22})
describes a system, the states of which are embedded partially in each
of the two different environments. These two different environments
are also independent of one another. They are, however, nothing but
parts of the total environment. 

Without singularity in the considered parameter range in relation to both
channels, we have   $E_i^{(1)} 
\approx E_i^{(2)}$, $\Gamma_i^{(1)} \approx \Gamma_i^{(2)}$ and 
$\Phi_i^{(1)} \approx \Phi_i^{(2)}$.
This case is realized in the decay of nuclear states since their decay 
probability can be determined  only when the decaying state is 
well isolated from other states. Otherwise, new problems arise from
the  overlapping with other states \cite{klro,ro91}.  
The above case is realized
also in the transmission when the individual transmission peaks are
well isolated from one another.

Under the influence of a singularity relative to $c=1$ and/or relative to
$c=2$, the eigenvalues and eigenfunctions will be, however, different
from one another,
$E_i^{(1)} \ne E_i^{(2)}$,  $\Gamma_i^{(1)} \ne \Gamma_i^{(2)}$ and 
$\Phi_i^{(1)} \ne \Phi_i^{(2)}$
in the corresponding parameter range.
Such a situation may occur in the transmission through a
quantum dot or quantum billiard \cite{burosa1,burosa2}.

In analogy to (\ref{ham0}), we will consider also
the non-Hermitian Hamiltonian
\begin{eqnarray}
\label{ham00}
{\cal H}_0^{(2,2)} = 
\left( \begin{array}{cccc}
\varepsilon_{1}^{(1)} 
 &
0 & 0 & 0  \\
0 & ~~~~\varepsilon_{2}^{(1)}  
& 0 & 0 \\
0 & 0 & \varepsilon_{1}^{(2)}  
 &
0  \\
0 & 0 &
0 & ~~~~\varepsilon_{2}^{(2)}   
 \\
\end{array} \right) 
\end{eqnarray}
which describes the system without any
mixing of its states via any environment. In other words, $\omega^{(1)}
=  \omega^{(2)}=0$ corresponds to vanishing EM of the eigenstates via
an environment.
The mixing of the eigenstates of (\ref{ham22}) can be represented in a
set of eigenfunctions of (\ref{ham00})
in complete analogy to the relation (\ref{eif12}) for two states
coupled to one common environment.

\subsection{Schr\"odinger equation with ${\cal H}^{(2,2)}$}
\label{sour2}

Using (\ref{ham00}), we can write down the Schr\"odinger equation with 
source term for the two-channel case in analogy to  (\ref{eif11})
for the one-channel case. The corresponding equation reads
\begin{eqnarray}
\label{eif11s4}
({\cal H}^{(2,2)}_0  - {\cal E}_i^{(c)}) ~| \Phi_i^{(c)} \rangle  = -
\left(
\begin{array}{cccc}
0 & \omega^{(1)} & 0 & 0\\
\omega^{(1)} & 0 & 0 & 0\\
0 & 0 & 0 & \omega^{(2)}\\
0 & 0 & \omega^{(2)} & 0
\end{array} \right) |\Phi_i^{(c)} \rangle \; . 
\end{eqnarray}
The source term depends on the coupling of   the system to both
channels, i.e. on  $\omega^{(1)}$ and  on $\omega^{(2)}$.
It does not depend on the energies $\varepsilon_{i}^{(c)}$.

We repeat here that, according to their definition \cite{kato},
EPs occur only in the one-channel case, i.e. only 
in the two submatrices related to channel $1$ and channel $2$, 
respectively.  They are not defined in the $4\times 4$ matrix 
(\ref{ham22}). 
 However, the EPs of the two submatrices in
 (\ref{ham22}) may influence the dynamics of the open two-channel 
 system.

\subsection{Non-Hermitian Hamiltonian and resonance structure 
of the  $S$ matrix}
\label{smatr}

An  expression for the $S$ matrix is derived 
and discussed in detail in \cite{ro03} by
rewriting the Breit-Wigner expression for one or more isolated
resonances. According to this derivation, the 
resonance structure of the $S$ matrix containing two resonance
states, can be obtained from the expression 
\begin{eqnarray}
\label{sm4}
S = \frac{(E-E_1-\frac{i}{2}\Gamma_1)~(E-E_2-\frac{i}{2}\Gamma_2)}{(E-E_1+
\frac{i}{2}\Gamma_1)~(E-E_2+\frac{i}{2}\Gamma_2)}
\end{eqnarray}
for  the case that two resonance
states are coupled to a common continuum of scattering
wavefunctions, see also \cite{top,proj10}. The expression (\ref{sm4}) is 
unitary. According to (\ref{sm4}), the resonance structure of the 
cross section is determined exclusively by the spectroscopic values 
of the localized part of the system, i.e. by 
the eigenvalues ${\cal{E}}_i = E_i + i/2~\Gamma_i$ of the
non-Hermitian  Hamiltonian $\cal H$. The expression  (\ref{sm4})
allows us therefore to obtain reliable results  
for the two channels   when the phase rigidity of the eigenfunctions 
of $\cal H$ is reduced ($r_k < 1$) and when the
eigenfunctions of  $\cal H$ contain EM, i.e. when they 
are mixed in the set of eigenfunctions 
$\{\Phi_i^0\}$ of ${\cal H}_0$ according to (\ref{eif12}).

As shown in \cite{top,proj10}, the $S$ matrix contains,
generally, nonlinear effects
which are caused by the EM of the resonance states via the environment. 
The one-channel case does, however, not allow us to prove the existence of
these nonlinear effects and of EM, since the resonance 
structure of the cross section calculated with and without EM 
is the same in this case \cite{proj10}. This result agrees, on the one
hand,  with the experience
received from many different numerical studies in realistic cases
which are performed  without taking into account EM. On the other hand,
it is not in contradiction with the conclusions received from the
study of non-Hermitian  physics of open quantum systems for the
following reason.

According to the results obtained in
\cite{proj10} for the one-channel case, the evolution of the system 
near to an EP is driven exclusively by the nonlinear source term of
the Schr\"odinger 
equation (\ref{eif11}) which describes the coupling of the localized 
part of the system to the common environment and is
characteristic of  the open quantum system embedded in one environment.
The calculations in \cite{proj10} are performed without varying  
$\omega$, i.e. $\omega$ can  not be responsible for the  
width bifurcation occurring in these calculations under the 
influence of an EP. Obviously, the nonlinear source term is 
the driving force. It is able, in the one-channel case, to largely 
conserve the resonance structure of the cross section. 

The conservation of the
resonance structure of the cross section which is
possible in the one-channel case, is expected to be impossible,
generally,  in the two-channel (or more-channel) case
according to (\ref{eif11s4}) since, generally, $\omega^{(1)} \ne
\omega^{(2)} \ne 0$. This gives the chance to 
test non-Hermitian quantum physics by means of the
two-channel case. We will provide
numerical results  and discuss their physical meaning in the following 
sections of the present paper.

\section{Numerical results}
\label{num}

\subsection{Eigenvalues and eigenfunctions of 
	${\cal H}^{(2,2)}$}
\label{spect}

\begin{figure}[ht]
\begin{center}
\includegraphics[width=12.cm,height=16.cm]{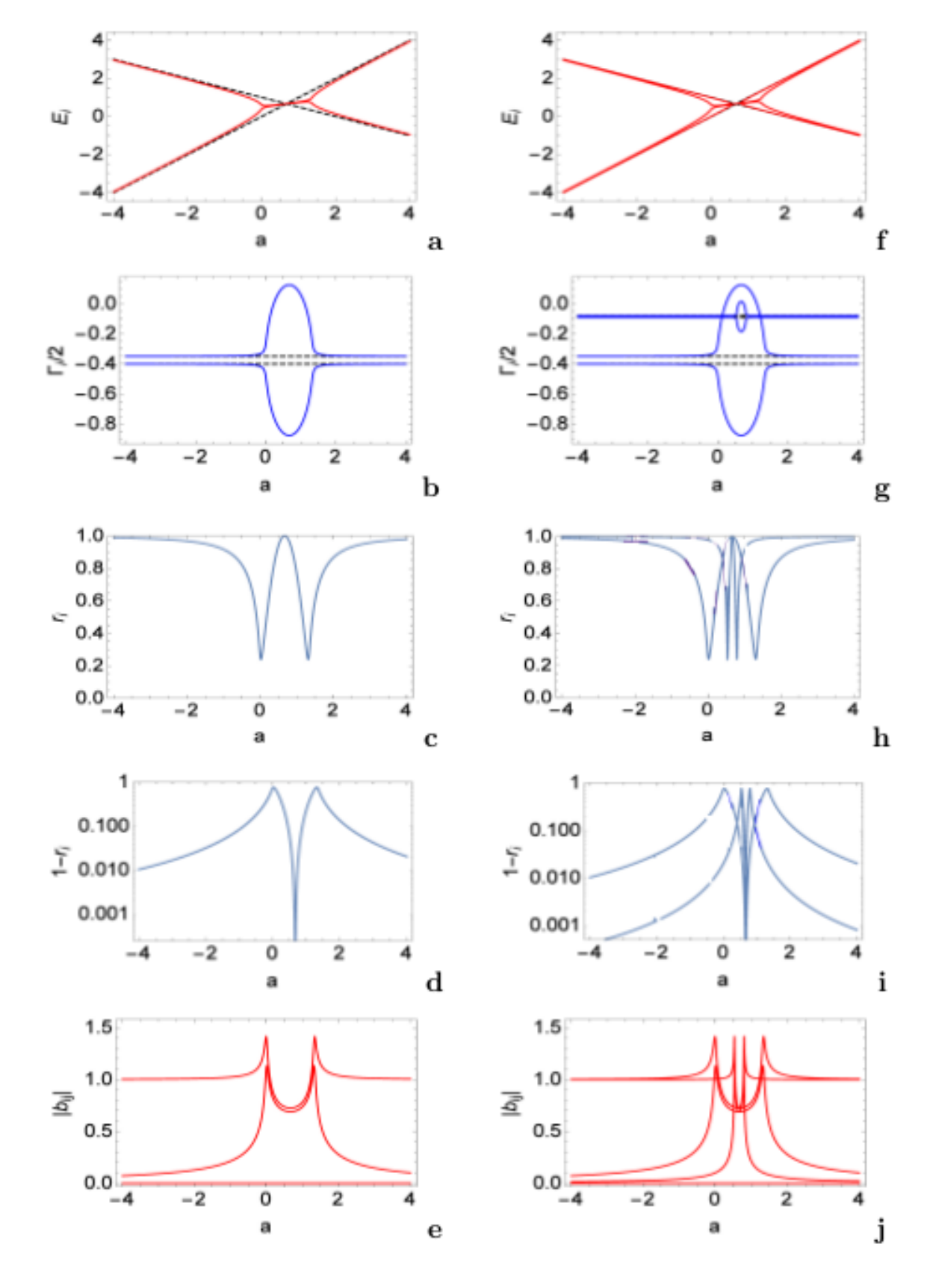}
\vspace*{-.7cm} 
\end{center}
\caption{
\footnotesize{
Eigenvalues ${\cal E}_i^{(1,2)} \equiv E_i^{(1,2)} +
\frac{i}{2}\Gamma_i^{(1,2)}$
and eigenfunctions $\Phi_i^{(1,2)}$
of the Hamiltonian ${\cal H}^{(2,2)}$ 
as a function of $a$.
 ~$\omega^{(1)} = \omega^{(2)} = 0.5i$ (left), 
~$\omega^{(1)} = 0.5i;  ~\omega^{(2)} = 0.1i$ (right).
 Parameters: $e_1= 1-a/2; ~e_2=a;$ 
left:
$\gamma_1^{(1)}/2= -0.4; ~\gamma_1^{(2)}=-0.35;
~\gamma_2^{(1)}/2=-0.35;
~\gamma_2^{(2)}/2=-0.4$ (dashed lines in a, b);
right:
$~\gamma_1^{(1)}/2= -0.4; ~\gamma_1^{(2)}=-0.08;
~\gamma_2^{(1)}/2=-0.35;
~\gamma_2^{(2)}/2=-0.09$ (dashed lines in f, g).
At the critical parameter value  $a=a^{\rm cr} = 0.6494$, 
the phase rigidity $r_i$ approaches the value $1$. 
}} 
\label{fig1}
\end{figure}

For illustration, we show in Fig. \ref{fig1}  a few
typical numerical results for two states in
the two-channel case. The results of the left column are obtained 
for the special case  $\omega^{(1)} =
\omega^{(2)}$ and those of the right column for the general case 
$\omega^{(1)} \ne \omega^{(2)}$. 
The numerical results for the eigenvalue trajectories 
${\cal E}_i^{(1,2)} \equiv E_i^{(1,2)} +
\frac{i}{2}\Gamma_i^{(1,2)}$
and the eigenfunction trajectories $\Phi_i^{(1,2)}$
of the Hamiltonian ${\cal H}^{(2,2)}$ are obtained
by starting from   parameter-dependent energies $e_i^{(1,2)}$  
and parameter-independent widths $\gamma_i^{(1,2)}$.
We consider  two different cases: in one case, the widths
$\gamma_i^{(1,2)}$ of the two states are equal,
$\gamma_i^{(1)}=\gamma_i^{(2)}$ (correspondingly to $\omega^{(1)} =
\omega^{(2)}$), while they are different from one another 
in the other case, $\gamma_i^{(1)} \ne \gamma_i^{(2)}$. 

The eigenvalue trajectories 
Figs. \ref{fig1}.a,b and   \ref{fig1}.f,g, respectively, 
show that the widths 
$\Gamma_i^{(1,2)}$ bifurcate in the neighborhood of an EP.
The energies of the two states are equal in this 
parameter range. At the critical
parameter value $a=a^{\rm cr}$  width bifurcation is maximum.
Here the phase rigidity approaches the value $1$
(Figs. \ref{fig1}.d and   \ref{fig1}.i),
meaning that the  two states become  orthogonal at this parameter
value. The EM of the states via the continuum can not be 
neglected at this parameter value 
(Figs. \ref{fig1}.e and   \ref{fig1}.j).

While the eigenvalue trajectories of  	${\cal H}^{(2,2)}$
are influenced by the critical point $a=a^{\rm cr}$
mainly in its very neighborhood, the 
eigenfunction trajectories are influenced in a 
comparably large parameter range. 
This can be seen, above all, in the phase rigidity
(Figs. \ref{fig1}.c,d,h,i) which is a quantitative measure for  the
biorthogonality of the eigenfunctions of 	${\cal H}^{(2,2)}$.
This fact is known also from calculations for the one-channel case 
\cite{proj10}.

We underline here once more that the results shown in Fig. \ref{fig1} 
are obtained for $N=2$ resonance states. 
They show therefore boundary effects arising from the parameter range 
in which no resonance states exist. 
In realistic systems, these boundary effects do, of course, not exist.   
Instead, we see an interference picture to which all $N$ states 
contribute, see e.g. \cite{klro,ro91}.

\subsection{Resonance structure and contour plot of transmission}
\label{cross}

\begin{figure}[ht]
\begin{center}
\includegraphics[width=12.cm,height=16cm]{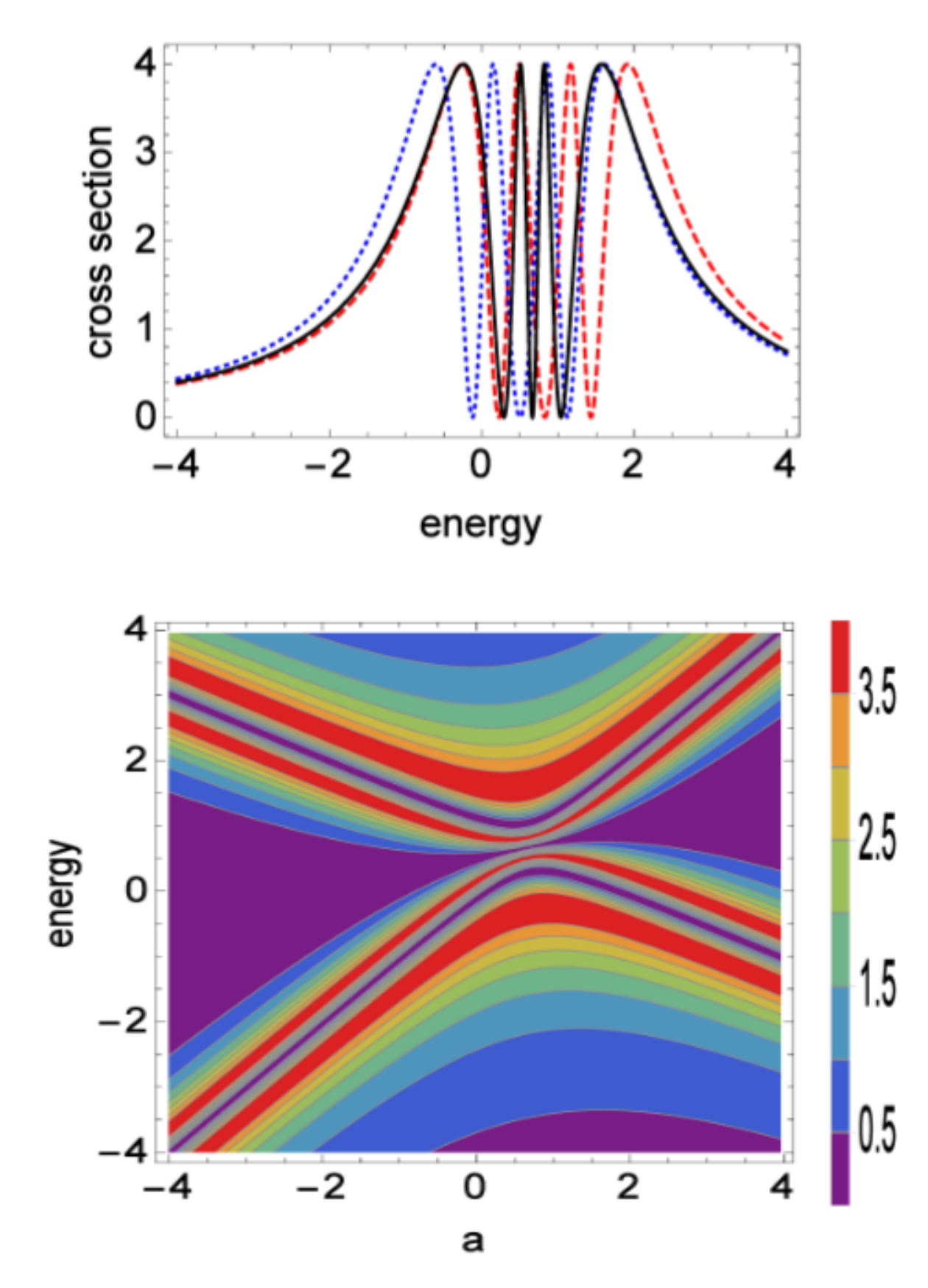} 
\vspace*{-.7cm}
\end{center}
\caption{
\footnotesize{
Resonance structure (above) and contour plot (below)
of the transmission with vanishing external mixing 
$\omega^{(1)}=\omega^{(2)} = 0$.   The
parameters are the same as in Fig. \ref{fig1} left.
~Above: full black line: $a=a^{\rm cr}=0.6949$; ~dashed red
line: $a_1= 1.3$; ~dotted blue line: $a_2= 0.0$. 
}	}
\label{fig2}
\end{figure}

\begin{figure}[ht]
	\begin{center}
\vspace*{1cm}
\includegraphics[width=12.cm,height=16cm]{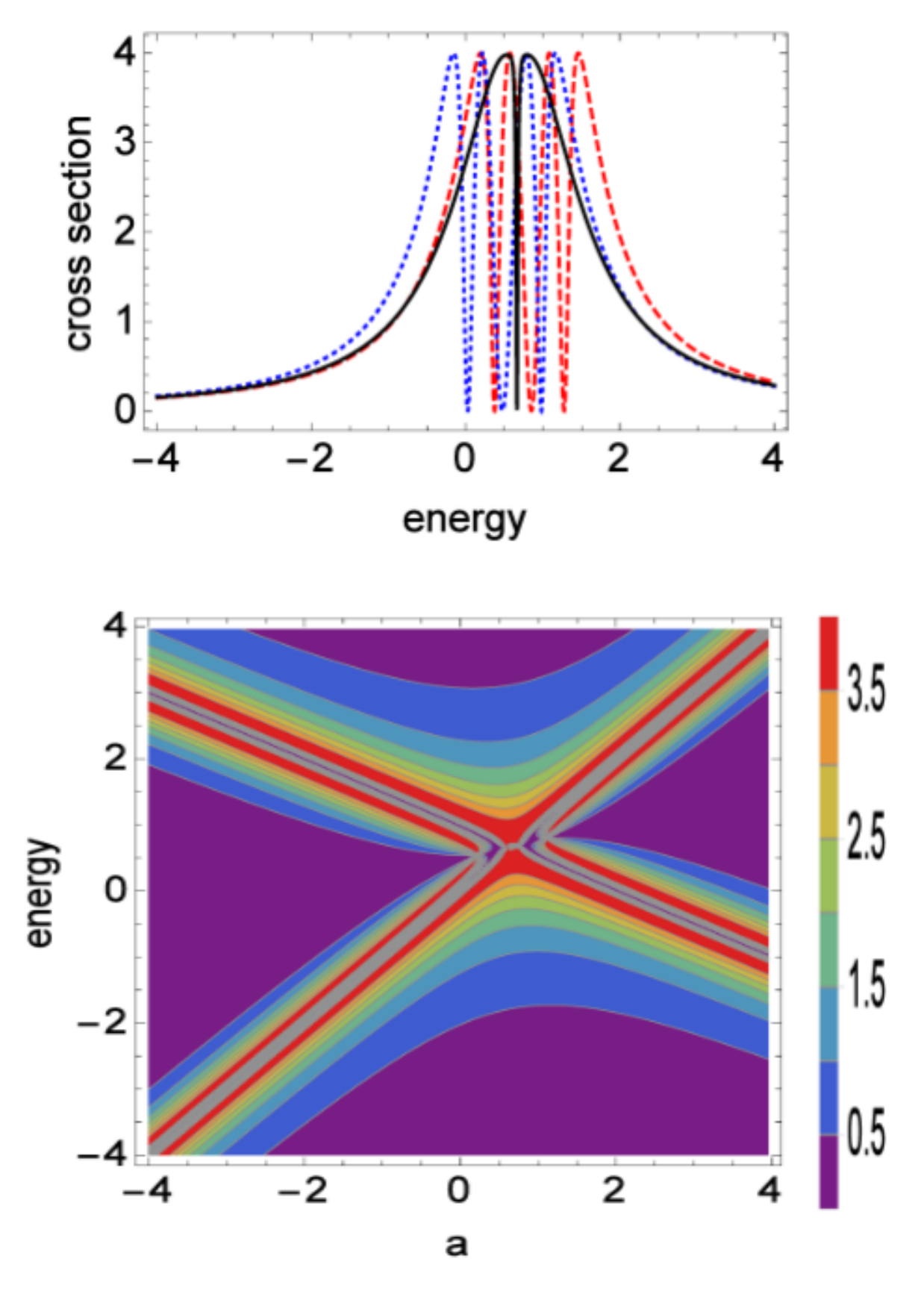} 
\vspace*{-.7cm}
\end{center}
\caption{
\footnotesize{The same as Fig. \ref{fig2} but  
$\omega^{(1)} \gg \omega^{(2)} \ne 0$. The parameters are the same as in
Fig. \ref{fig1} right.
}}
\label{fig3}

\end{figure}

\begin{figure}[ht]
	\begin{center}
\vspace*{1cm}
\includegraphics[width=12.cm,height=16cm]{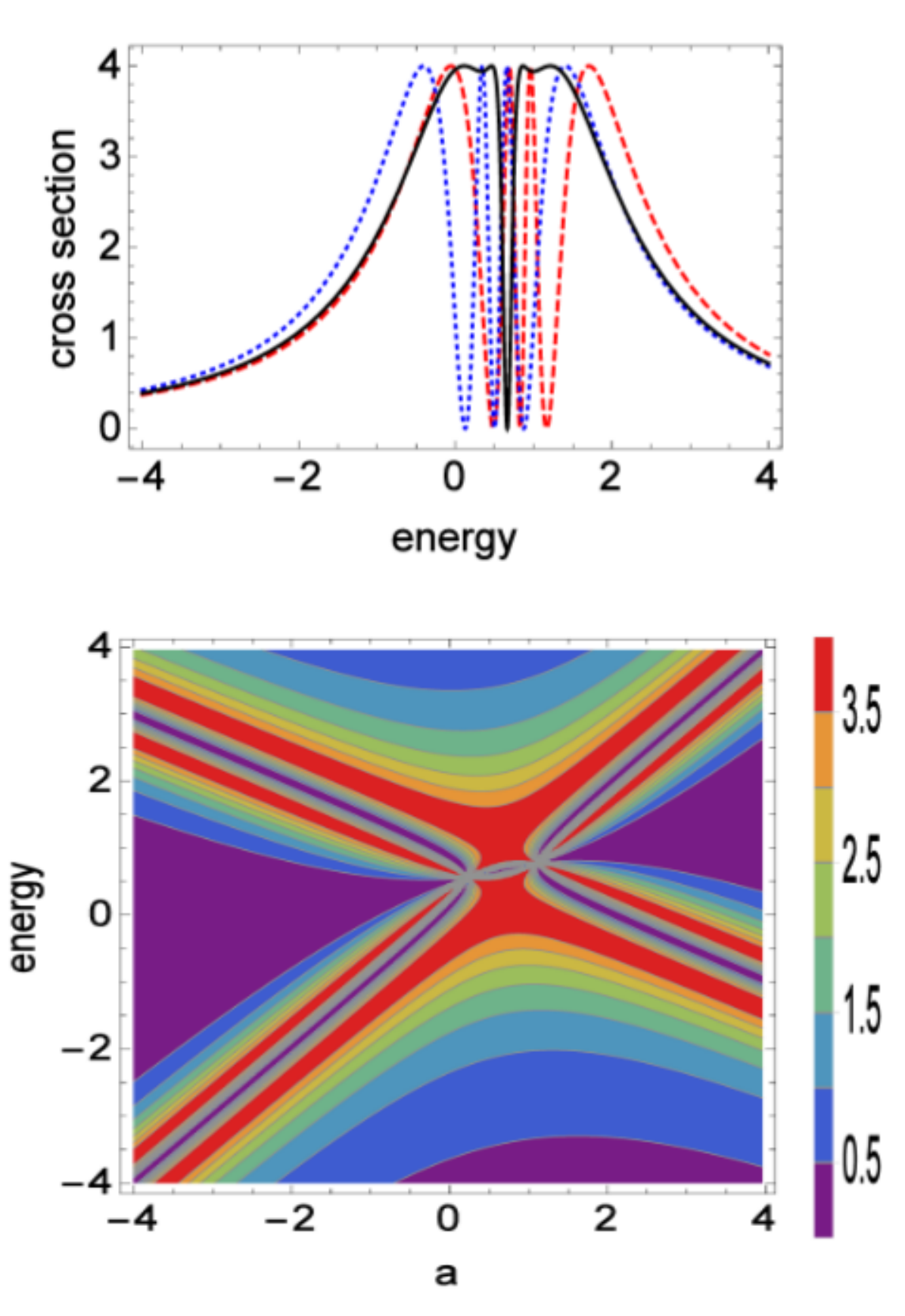} 
\vspace*{-.7cm}
\end{center}
\caption{
\footnotesize{
The same as Fig. \ref{fig2} but $\omega^{(1)} = \omega^{(2)} \ne 0$.
 The parameters are the same as in Fig. \ref{fig1} left
}}
\label{fig4}
\end{figure}

Interesting information on the spectroscopic properties of 
the localized part of the system  
is contained in the resonance structure of the cross
section which can be observed experimentally.
We consider the  transition from
channel $1$ to channel $2$ which simulates 
transmission. Using (\ref{sm4}), we have calculated 
the resonance structure of the transmission
at the critical parameter value $a^{\rm cr}$ and at two values
of $a$ beyond the critical  range (above) as well as the
corresponding contour plots (below).
 
In Fig. \ref{fig2}, the results without EM (corresponding to
$\omega^{(1)} =  \omega^{(2)} = 0 $) are shown.
There appear four states, indeed, according to the results of
analytical studies. This picture is different from 
the resonance structure of the cross section 
of a two-level system coupled to  one
channel characteristic of which is its double-hump structure   
around maximum width bifurcation \cite{top,proj10}.

Figs. \ref{fig3} and \ref{fig4} show the transmission 
through a two-level system coupled to  two channels 
with $\omega^{(c)} \ne 0$   in the 
same parameter range around  maximum width bifurcation.
These figures illustrate that the double-hump structure  
of the cross section is restored  in the transmission 
(i.e. in the two-channel case) under the influence of EM.  

In Fig. \ref{fig3} (with different coupling strengths 
$\omega^{(1)} \gg \omega^{(2)}$ according to the eigenvalues and
eigenfunctions of Fig. \ref{fig1} right), 
the double-hump structure around the maximum width distribution
is very similar, indeed, to that of the cross section in the
one-channel case (solid black line in Fig. \ref{fig2} compared to
solid black line in Fig. 4.a in \cite{proj10}). The contour plot of 
the transmission  is however much richer than 
the cross section in the one-channel case. This is reflected also in
the dotted (blue) and  dashed (red) lines of Fig. \ref{fig3}.a.

More interesting are the results shown in Fig. \ref{fig4}
corresponding to the eigenvalues and eigenfunctions of Fig. \ref{fig1}
left with $\omega^{(1)} = \omega^{(2)}$.
Here, the double-hump structure also appears. The bumps are however 
broadened in energy. Due to this broadening, the transmission 
is enhanced in a comparably large energy window.  In this energy 
window, the phase rigidity averaged over energy,
is reduced (see Fig. \ref{fig1}).

\section{Discussion of the results}
\label{disc}

\subsection{External mixing}
\label{disc1}

The results shown in Figs. \ref{fig1} to \ref{fig4} show very clearly 
that the mixing of the states via the environment
(EM) plays an important role in open
quantum systems. This mixing is a second-order 
effect and does not appear in standard Hermitian quantum physics. 
While its influence onto the spectroscopic properties of the 
localized part of the system far from singular points 
is usually small, 
it becomes important and determines the dynamics of the system 
in a finite neighborhood of the critical parameter values $a=a^{\rm cr}$.

\subsection{Meaning of "points"}
\label{disc2}

Our results show further that the dynamical
properties of the system are influenced by EPs not only at 
parameter values that correspond to their "exact"
position. Almost the same properties arise 
near to these parameter values. This can be
seen, e.g., in the width trajectories in Figs. \ref{fig1}.b and 
\ref{fig1}.g  which bifurcate although the parameter
values do not allow an exact crossing point of the trajectories.

\subsection{Two-channel systems and EPs}
\label{disc3}

According to the mathematical studies by Kato \cite{kato}, EPs are 
defined in relation to  one  continuum. Their physical 
meaning is studied therefore, up to now, mostly in systems that are 
embedded in one well-defined common environment. 
For references see the review \cite{top}.

Most states of physical systems are coupled, however, to more than one 
channel. For example, the transmission through
a small system (e.g. a quantum dot) is related to, at least, two
channels, the entrance and the exit channel. The transmission
of particles through such systems 
is studied by many authors in many different papers by using 
different methods. The most popular methods start from a Hermitian
Hamilton operator and consider parameter ranges beyond EPs.

\subsection{Transmission}
\label{disc4}

In the present paper, transmission is  studied 
for the first time  in the framework of the non-Hermitian quantum 
physics without additional assumptions. The only assumption is
that the (localized) system is embedded in more than one environment 
(meaning that it is coupled to more than one channel). 
Although the expression (\ref{ham22}) 
for the Hamiltonian ${\cal H}^{(2,2)}$ seems to be arbitrary
or a mathematical subtleness, the obtained
results  show that this genuine non-Hermitian
Hamiltonian describes the properties of the two-channel system 
according to expectations.

\subsection{One-channel versus two-channel systems}
\label{disc5}

We compare the results obtained in the present paper
for the two-channel system, with those known for open quantum 
systems embedded in one common environment. 
The following properties are independent of the number of channels. 

(i) Far from singular points, the
spectroscopic properties of the system are influenced 
only marginal by its embedding into an environment. 
They are, however, never exactly the same as those of a closed system.  
This fact is proven experimentally \cite{savin2}.

(ii) Singular points change the system properties in a certain finite
neighborhood of their exact position. Here, they cause the same observable
effects which are expected at their ``exact'' position.  

(iii) At the critical parameter value  $a^{\rm cr}$, the
eigenfunctions of the non-Hermitian Hamilton operator $\cal H$ are
orthogonal (and not biorthogonal). At this parameter value, width
bifurcation is maximum; and, in the two-channel case, the transmission
through the system is
enhanced. This  effect is observable.

The one-channel case has however some special features. 
The nonlinear processes involved in the non-Hermitian dynamics, 
restore the characteristic resonance structure of the
cross section that has been 
obtained without taking into account EPs and  EM.
It is possible therefore to describe the system 
properties, in the one-channel case, without taking into
account the characteristic features of an open quantum system  
\cite{proj10}. The figures \ref{fig1} to \ref{fig4}
of our paper show clearly that this possibility is really restricted 
to the one-channel case. The information involved in the  
more-channel cases is much richer.

\subsection{Coherence in two-channel systems}
\label{disc6}

In the  two-channel case EPs cannot be seen. 
We see only the critical value  $a^{\rm cr}$ of the parameter $a$
which gives rise to new observable effects such as the enhancement of 
the transmission at maximum width bifurcation. 
This effect is measurable.
The phase rigidity is $1$ at
$a^{\rm cr}$ because  the eigenfunctions of $\cal H$ are
orthogonal at this point. It is however
$r_k < 1$ in the neighborhood of  $a^{\rm cr}$.
Averaged over a certain parameter range around  $a^{\rm cr}$, the
phase rigidity is
therefore reduced, while the transmission is enhanced.

In \cite{burosa2}, the transmission through a system with many states 
is calculated  numerically by using the tight-binding lattice model
(according to Datta \cite{datta}) 
with one channel in each of the two attached identical leads. As a
result of this calculation,  the  phase rigidity, averaged
over the considered energy window, and the transmission are 
anti-correlated. Thus, the two very different methods (each of which
solves the problem exactly) produce qualitatively the same results.

These results show  that 
coherence appears in two-channel (and more-channel) systems
in competition with dissipation. Coherence is known to play an
important role in
 biophysical systems. According to our results, it is characteristic
 of all open quantum
 systems which are coupled to more than one open channel. 
It may be small; it is however nonvanishing.

We underline once more that the above mentioned processes are related 
exclusively to the properties of the  eigenstates of a non-Hermitian
Hamilton operator.
They are  not involved in any version of Hermitian quantum physics.
For example,  EM can be simulated, to some extent, in the
standard calculations with Hermitian Hamiltonian by including it
effectively into the Hamiltonian. Its interesting relation to the 
critical parameter value  $a^{\rm cr}$ can, however, not be seen in 
such calculations.

\section{Concluding remarks} 
\label{concl} 

Our results show that the genuine 
non-Hermitian Hamiltonian $\cal H$ can be used  for the description
of the characteristic features 
of many-body systems with transfer of particles. In this case,
the Hamiltonian $\cal H$ is surely not the most convenient  one
in order to describe a realistic system. It demonstrates however that 
non-Hermitian quantum physics is a powerful method that
can explain many different features of open quantum systems,
including those that can be described successfully in standard theory. 

It is very well known that the non-Hermitian part of the Hamiltonian causes 
dissipation, which is characteristic of all open quantum systems. 
In competition with dissipation, coherence appears when the system is
coupled to at least two open channels. The
competition between coherence and dissipation becomes the more
important the more channels are open. It is the most interesting
feature of realistic open quantum systems. 

The results of the present paper show furthermore that
the eigenfunctions of the non-Hermitian Hamilton operator $\cal H$
play  an important role, above all their non-rigid phases
around the singular EPs; and their non-vanishing external mixing
via the environment. 
At the critical parameter value $a^{\rm cr}$, a short-lived state 
results  from width bifurcation. 
This state is similar to the so-called superradiant state discussed
in, e.g., \cite{cel1,cel2}. The mechanism of the formation 
of the superradiant state is however
completely different from that of the short-lived state 
discussed in the present paper.

The critical point $a^{\rm cr}$ at which the eigenfunctions of the
non-Hermitian Hamilton
operator $\cal H$ are orthogonal (and not biorthogonal), 
appears in the one-channel  as well as
in the two-channel problem (for the one-channel 
case see \cite{proj10}). Here, the  quantum system and 
the environment into which it is embedded, are in a
state of equilibrium \cite{nh7}.   

The same formalism of non-Hermitian quantum
physics that is able to describe  properties of many-particle systems,
can explain also  experimental results
which are puzzling in standard Hermitian quantum physics.
Examples are the explanation \cite{muro} of the so-called phase lapses 
in mesoscopic systems as well as the 
understanding \cite{proj12} for the high efficiency of photosynthesis. 

Summarizing we state that open quantum systems embedded in two 
(or more) environments
are abounded in very many interesting aspects of non-Hermitian 
quantum physics. 
They allow, on the one hand, to study characteristic
features of non-Hermitian quantum physics. 
On the other hand, they  provide the possibility of
many promising applications.

\vspace{0.8cm}

{\bf Acknowledgment}

We are indebted to Jon Bird for valuable discussions.


\begin{thebibliography}{99}
\bibitem{comment1} In contrast to the definition that is used in, 
for example, nuclear physics, we define the complex energies 
before and after diagonalization of ${\cal H}$ by 
$\varepsilon_k = e_k + \frac{i}{2} \gamma_k$ and  
${\cal E}_k = E_k + \frac{i}{2} \Gamma_k$, respectively, 
with $\gamma_k \le 0$ and $\Gamma_k \le 0$ for decaying states. 
This definition will be useful when discussing systems with  gain
(positive widths) and  loss (negative widths).
\end{thebibliography}
\end{document}